\documentclass{article}
\usepackage[T1]{fontenc} 
\usepackage[utf8]{inputenc} 
\usepackage{ismir,amsmath,cite,url}
\usepackage{graphicx}
\usepackage{color}

\usepackage{caption}
\usepackage{subcaption}
\usepackage{bm}
\usepackage{amssymb}
\usepackage{dsfont}
\usepackage[linesnumbered,ruled,vlined]{algorithm2e}
\usepackage{csquotes}
\usepackage{verbatim} 
\usepackage{minibox}
\usepackage{tikz}
\usepackage{float}
\usepackage{stfloats}

\makeatletter
\newcommand*{\radiobutton}{%
  \@ifstar{\@radiobutton0}{\@radiobutton1}%
}
\newcommand*{\@radiobutton}[1]{%
  \begin{tikzpicture}
    \pgfmathsetlengthmacro\radius{height("X")/2}
    \draw[radius=\radius] circle;
    \ifcase#1 \fill[radius=.6*\radius] circle;\fi
  \end{tikzpicture}%
}
\makeatother

\SetKwInput{KwInput}{Input}                
\SetKwInput{KwOutput}{Output}              

\usepackage{lineno}
\usepackage[bookmarks=false]{hyperref}

\title{Generating music with sentiment using Transformer-GANs}

\threeauthors
  {Pedro L. T. Neves} {State University of Campinas \\ {\tt p185770@dac.unicamp.br}}
  {Jose Fornari} {State University of Campinas \\ {\tt fornari@unicamp.br} }
  {João B. Florindo} {State University of Campinas \\ {\tt florindo@unicamp.br}}

\def\authorname{Pedro L. T. Neves, Jose Fornari, and João Batista Florindo}

\begin{document}

\maketitle

\begin{abstract}
The field of Automatic Music Generation has seen significant progress thanks to the advent of Deep Learning. However, most of these results have been produced by unconditional models, which lack the ability to interact with their users, not allowing them to guide the generative process in meaningful and practical ways. Moreover, synthesizing music that remains coherent across longer timescales while still capturing the local aspects that make it sound ``realistic'' or ``human-like'' is still challenging. This is due to the large computational requirements needed to work with long sequences of data, and also to limitations imposed by the training schemes that are often employed. In this paper, we propose a
generative model of symbolic music conditioned by data retrieved from human sentiment. The model is a Transformer-GAN trained with labels that correspond to different configurations of the valence and arousal dimensions that quantitatively represent human affective states. We try to tackle both of the problems above by employing an efficient linear version of Attention and using a Discriminator both as a tool to improve the overall quality of the generated music and its ability to follow the conditioning signals. 
\end{abstract}


\section{Introduction}

One of the driving factors behind the human experience of music is the emotional content that it conveys. Several works in the area of Musical Information Retrieval (MIR) have focused on Music emotion recognition, that is, automatic recognition of the perceived emotion of music based solely on the musical information itself \cite{music-emotion}. In this context, emotion is often represented according to the valence and arousal dimensions originated from to the Russell circumplex model \cite{valence-arousal}. While valence expresses how pleasurable or displeasurable an emotion is, arousal represents the level of alertness associated with that emotion, going from relaxed to excited. 
However, within the realm of \textit{Deep Learning} comparatively little research has focused on reverting this process, that is, instead of predicting the affective content of a musical passage, being able to generate musical pieces that project a desired emotional state. 

Deep Neural Networks are now capable of generating songs that display coherent chord progressions, melodies and even lyrics \cite{jukebox}, despite the fact that these characteristics often do not persist across longer time scales. While there are several reasons behind this fact, two of them stand out. Firstly, there are the inherent computational and memory costs involved in modeling longer sequences of data. Secondly, the most common technique used to train these models, teacher forcing or Maximum Likelihood Estimation (MLE) \cite{teacher-forcing}, makes them work with data distributions that are different during training and inference time (real vs synthetic). This is known as exposure bias \cite{exposure-bias}. One of the ways to alleviate this problem is by delegating the task of judging which samples are good and which are not to a different neural network that is trained in conjunction with the generative model. This is the motivation behind the use of Generative Adversarial Networks (GANs) \cite{gan} within the realm of sequence generation \cite{seqgan}.



In this work, we present a generative model of music capable of synthesizing songs conditioned by values of valence and arousal that correspond to perceived sentiment \cite{valence-arousal}. The ability of automatically generating music that follows a specific pattern of emotions can be interesting in various contexts, e.g. producing soundtracks to accompany story-driven forms of media such as Video-Games and Movies, which often use music as a means of guiding the audience towards a specific emotional state that suits the narrative. The goal here is to provide users who may not have the musical background necessary for composition a way of translating their perception into songs that can suit their artistic aspirations.

In an effort to mitigate the shortcomings of teacher forcing-style training, we complement it with an additional adversarial signal provided by a Discriminator network. This addition has a positive effect on the generated samples, and we demonstrate, through evaluations of our model both via automatic metrics and human feedback, that the proposed Transformer GAN obtains a performance that competes with a current state-of-the-art model, even while having a smaller set of parameters and using a simpler representation of music. To summarise, our contributions are as following:

\begin{itemize}
    \item We present a neural network that, up to our knowledge, is the first generative model based on GANs to produce symbolic music conditioned by sentiment.
    
    \item We show, both through automatic and human evaluation, that our model obtains a competitive performance with that of a current state-of-the-art model on the task of music generation conditioned by sentiment.
    
    \item We show that promising results come from using Generative Adversarial Networks within the context of music generation conditioned by sentiment, as our model obtains a good performance despite having less parameters, using a simpler symbolic representation, and, being, up to our knowledge, the first on this task to be trained via an adversarial scheme.
\end{itemize}

\section{Related Works}

\subsection{Generative Adversarial Networks}

Generative Adversarial Networks, or simply GANs, consist of a theoretical framework in which two Neural Networks, the Generator and the Discriminator, through competition, optimize a model that implicitly approximates a data distribution by generating samples that try to mimic the features that it observes on a given set of samples originating from that distribution. Each of the networks is trained to optimize an objective function. The objective of the Discriminator $D$ is to separate the real samples from those that are created by the Generator $G$, whose job is to produce samples that are so similar to those of the real distribution that the Discriminator is unable to determine which of them are real and which are fake. This whole process is equivalent to a min-max game that can be formalized by Equation \ref{eq:gan}.

\begin{align}\label{eq:gan}
\begin{split}
    \min\limits_{G} \ \max\limits_{D} \ V(D, G) =  & \mathbb{E}_{\bm{x} \sim p_{data}(\bm{x})}[\log D(\bm{x})]+ \\ & \mathbb{E}_{\bm{z} \sim p_{z}(\bm{z})}[\log(1-D(G(\bm{z})))].
\end{split}
\end{align}
In the equation above, $p_{\text{data}}(\bm{x})$ is the real data distribution, and $p_z(\bm{z})$ is a prior on input noise variables that come from a normal distribution, and which are then mapped to data space by $G$, so that the synthetic distribution, $p_g$, can be learned.


Due to the inherent instability of the adversarial process, several works have focused on improving the convergence and the quality of the samples generated by GANs via new objective functions, regularization and normalization techniques, and model architectures. Of particular interest to this work is RSGAN \cite{rsgan}, which substitutes the standard GAN loss for the non-saturating Relativistic Standard Loss. Here, as the authors put it, the Discriminator estimates the probability that the given real data is more realistic than a randomly sampled fake data. Equations \ref{eq:rsgan_d} and \ref{eq:rsgan_g} correspond to the objectives for the Discriminator and the Generator, respectively.

\begin{align}\label{eq:rsgan_d}
\begin{split}
   &\mathcal{L}_{RSGAN, D} = \\ &-\mathds{E}_{(x_r, x_f)\sim (\mathds{P}, \mathds{Q})}[\log(\text{sigmoid}(D(x_r) - D(x_f)))]
\end{split}
\end{align}

\begin{equation}\label{eq:rsgan_g}
\begin{split}
    &\mathcal{L}_{RSGAN, G} = \\&-\mathds{E}_{(x_r, x_f)\sim (\mathds{P}, \mathds{Q})}[\log(\text{sigmoid}(D(x_f) - D(x_r)))],
\end{split}
\end{equation}
in which $\mathds{P}$ and $\mathds{Q}$ are, in this order, the real and fake disributions.




 In order to provide more stability to the training process, WGAN-GP \cite{wgan-gp} introduces a Gradient Penalty in the form of an additional loss that enforces a Lipschitz constraint on the Discriminator. This loss is expressed in Equation \ref{eq:wgan_gp}:

\begin{equation}\label{eq:wgan_gp}
   \mathcal{L}_{GP} = \mathds{E}_{\hat{x} \sim \mathds{P}_{\hat{x}}}[(\|\nabla_{\hat{x}} D_{\phi}(\hat{x}) \|_2 - 1)^{2}]
\end{equation}
where $\hat{x}$ is sampled along straight lines between pairs of points taken from the real and fake data distributions, and $\phi$ are the Discriminator parameters.





One possible strategy that can be used to combat the effect known as exposure bias \cite{exposure-bias}, characterized by the disparity between the data available to the network during training and inference time when the model is trained via standard Maximum Likelihood Estimation (MLE), is the insertion of a Discriminator into the training process as a tool to guide the Generator. Nevertheless, generating discrete sequences with GANs is notoriously hard. This is mostly due to the fact that the outputs of some sequence models are discretized in a way that prohibits the gradient of the Discriminator loss to propagate through the Generator. Several strategies have been proposed to circumvent this issue \cite{reinforce, gumbel-softmax, gumbel-softmax-gan}. Here, we employ the Gumbel-Softmax technique presented in \cite{gumbel-softmax} and applied to adversarial training in \cite{gumbel-softmax-gan}. Mathematically, if we have a categorical distribution with class probabilities $\pi_i, i\in{i,...,d}$, we can draw samples y from this distribution using: 

\begin{equation}\label{eq:gumbel-sample}
        y = \text{one$\_$hot}\left(\text{arg}\max_{i}[g_i +\log \pi_i]\right)
\end{equation}
where each $g_i, i\in{i,...,d}$ is taken from a Gumbel Distribution with location $0$ and scale $1$. From this expression, one can obtain a continuously differentiable approximation of the categorical distribution parameterized in terms of the softmax function:

\begin{equation}\label{eq:gumbel-softmax}
        y = \text{softmax}((1/\tau) (\pi + g)),
\end{equation}
where $\tau$ is a temperature parameter that regulates how close to the categorical (lower $\tau$) \textit{versus} to the uniform distribution (higher $\tau$) is $y$. This parameter is annealed from large values to close to zero during training. Here, we use the same schedule as in \cite{transformer-gan2}, that is, $1/\tau = (1/\tau_{\min})^{n/N}$, where $n$ is the index of the current global optimization, $N$ is the total number of steps and $\tau_{\min}$ is a hyperparameter which we chose to be $10^{-2}$.

\subsection{Transformers}

Shortly after its presentation in 2017, the Transformer \cite{transformers} became the most popular architecture in the field of Natural Language Processing (NLP), and it has also been successfully applied to other areas, such as image recognition \cite{vision-transformer} and audio \cite{transformers-audio}. The main reason behind its success is the Self-Attention mechanism, which allows it to model relationships between all elements of a sequence. The mechanism takes in matrices $Q$, $K$, and V, corresponding, respectively, to Queries, Keys and Values, and calculates a weighted sum of the values where the coefficients are given by a similarity function between the queries and the keys.




One of the models developed with the intent of decreasing the computational costs associated with the calculation of Attention matrices \cite{sparse-transformers, linear-transformer, reformer, performer} is the  Linear Transformer \cite{linear-transformer}, which, as the name suggests, employs a version of the attention mechanism with computational and memory requirements that grow linearly with respect to sequence length (standard Attention has quadratic requirements), and is also faster than regular attention. This is achieved via the substitution of the standard softmax similarity score by one that allows the matrix multiplications necessary for the calculation of the attention matrix to be factorized in a more efficient way. If that kernel has a feature representation $\phi(x)$, one can write the Attention matrix as:

\begin{equation}\label{eq:lin-attention}
    A(Q, K, V)_i = \frac{\sum_{j=1}^{N}\phi(Q_i)^T\phi(K_j)V_j}{\sum_{j=1}^{N}\phi(Q_i)^T\phi(K_j)}.
\end{equation}

With this expression, it is possible to calculate the factor inside the sum only once, and to reuse it to find all the queries. Specifically, the authors use the kernel with a feature map that results in a positive similarity function: $\phi(x) = \text{elu}(x) + 1$, where elu is the exponential linear function \cite{elu}. In this same work, the authors also discover an analogy between the Transformer and the RNN. We refer to \cite{linear-transformer} for further details on the Linear Transformer.

\subsection{Generative Models of Music}




As of the writing of this paper, most of the state-of-the art generative models of symbolic music are Transformers or Transformer-based. Music Transformer \cite{music-transformer} uses relative self-attention \cite{relative-attention} to process longer sequences of data and generate music that exhibits long-term structure. MuseNet \cite{musenet} is able to produce multi-track compositions spanning several musical genres and artists by employing time, note and structural embeddings that give the model more context.

Midinet \cite{midinet} and MuseGAN \cite{musegan} are GAN-based generative models of music that use Convolutional Neural Networks (CNNs) as both the Generator and Discriminator. In those works, music is represented in the form of piano-rolls that work analogously to images. Most similar to our work are \cite{transformer-gan1}, where the Discriminator is trained to judge the content both locally and globally, and \cite{transformer-gan2}, which uses a Transformers-XL \cite{transformer-xl} as Generator and a pre-trained BERT \cite{bert} as Discriminator, and also employs the Gumbel-Softmax trick discussed above.

Several works have also explored conditioning musical generation with emotional content. In \cite{sentimozart}, human emotions are captured from images of human faces and then categorized into 7 categories. Then, the researchers try to generate music based on these emotions. In \cite{emotion-lstm}, Biaxial LSTM networks are used to produce polyphonic music, and the generation can be conditioned by emotion via 4 parameters originating from the valence and arousal dimensions. In \cite{art-music}, the authors translate between the visual and musical domains via emotions. Specifically, they use neural networks to extract emotional content from images and music, and subsequently feed the content originated from both pieces of data into a network to condition music generation. Music Fadernets \cite{music-fadernets} constitute a framework that can infer high-level feature representations by first modelling their equivalent low-level attributes, which are easier to quantify. These low-level features are then used for style transfer across arousal states. In \cite{bardo-composer}, a model dubbed Bardo Composer is presented as a system to generate backgorund music for tabletop role-playing games. This is done through Stochastic Bi-Objective Beam Search (SBBS), a search algorithm that samples from a distribution of sequences and selects for one that maximizes for realism and emotion. Furthermore, \cite{generating-music-sentiment} uses a genetic algorithm to influence specific LSTM units that learn to encode sentiment in a pre-training language modeling stage, allowing it to steer the passages that it generates towards a desired affective state. In \cite{lead-sheets-affect}, the authors use pre-defined mood-tags associated with each chord in a progression to guide the generative process step-by-step. The EMOPIA dataset \cite{emopia} contains musical passages separated into four quadrants that each corresponds to a combination of positive or negative arousal and valence. The excerpts on the dataset are from piano transcriptions of pop songs that were labeled by its authors. In this same work, the authors also use a Transformer model that employs the Compound Word representation \cite{cp} to generate songs conditioned by affective states.

One of the factors that influence the final quality of the samples generated by models of symbolic music is the representation used to train them. For contemporary styles, like Pop or Hip-Hop, where a rigid metrical grid is often followed, it is desirable to incorporate data about the rhythmic structure of the songs into the representation. REMI \cite{remi}, which stands for revamped MIDI-derived events, is a beat-based approach to modeling music that encodes this information through tokens that signal the beginning of a bar and the passage of each beat, while still maintaining some flexibility by allowing local tempo changes. More specifically, there NOTE$\_$ON, NOTE$\_$DURATION, VELOCITY, TEMPO, BAR and BEAT. A NOTE$\_$ON event indicates the start of a note, NOTE$\_$DURATION corresponds to the duration of that note, and VELOCITY is a parameter that indicates the intensity with which the keys on a piano are played, and correlates to the volume of the note produced. Finally, TEMPO dictates the tempo of the musical excerpt from the moment the token is produced forward, and BAR events indicate the start of a new bar. The Compound-Word Transformer \cite{cp} uses a separate embedding for each type of musical element (pitch, chord, tempo value, etc.).



\section{Methods}

\subsection{Architecture and Loss functions}
Both the Generator and Discriminator are Transformers with linear versions of the Attention Mechanism \cite{linear-transformer}. 
Each of the models is composed by $6$ Attention blocks.

The Generator has two roles. In the first place, it has to predict each item of each sequence from the real dataset based on the previous elements, that is, it has to complete pieces of already existing sequences. In standard Transformer fashion, a lower triangular or look-ahead mask is applied to the original sequence in order to prevent the model from seeing its future elements. The second objective of the network is to generate sequences that are similar to those of the real set from scratch, that is, without context from the real dataset, such that these sequences can fool the Discriminator. These sequences are generated step-by-step in an autoregressive manner, which is done via the Transformer-RNN analogy made in \cite{linear-transformer}.
¨

 The Generator is conditioned via special scale and bias parameters that influence the Layer Normalization \cite{layernorm} layers existing within the Attention mechanism. There is one of these couples for each class on the dataset and one for the unlabeled sequences. Formally, if $i$, $k$ and $c$ stand, respectively, for position in the sequence, feature channel and class label, we have:

\begin{equation}\label{eq:cond_g}
   s_{i, k}^{\prime} = \gamma_{k}^{c}\cdot s_{i, k} + \beta_{k}^{c}
\end{equation}
where $s$ and $s^{\prime}$ are input and output sequences, and $\gamma$ and $\beta$ are scale and bias.

The Discriminator takes each sequence as a whole and tries to determine if it is real or fake. To design this network, we took inspiration from the Visual Transformer \cite{vision-transformer}, separating the sequences into patches of a certain length and transforming each patch into a single feature vector. 

Our Discriminator has two outputs. First, there is a single feature unit that indicates if the passage originates from the real or fake datasets and if it exhibits the desired characteristics provided by the conditional signal. To produce this output, a [CLS] token is concatenated to the input sequence, similarly to BERT \cite{bert}. Then, the conditional information is incorporated via an inner product between an embedding of this information and the [CLS] representation. This essentially means that the model works as a Projection Discriminator \cite{proj-disc}. The second output is a prediction map where each unit corresponds to a single patch in the sequence, that is, for every collection of musical symbols with length equal to the patch size, there is a value predicting whether that patch is real or fake. This technique, often used in image generating-GANs \cite{pix2pix}, ensures that the model prioritizes local structure. 

\begin{figure}[!ht]
    \centering
     \begin{subfigure}[b]{0.35\textwidth}
         \centering
         \includegraphics[width=\textwidth]{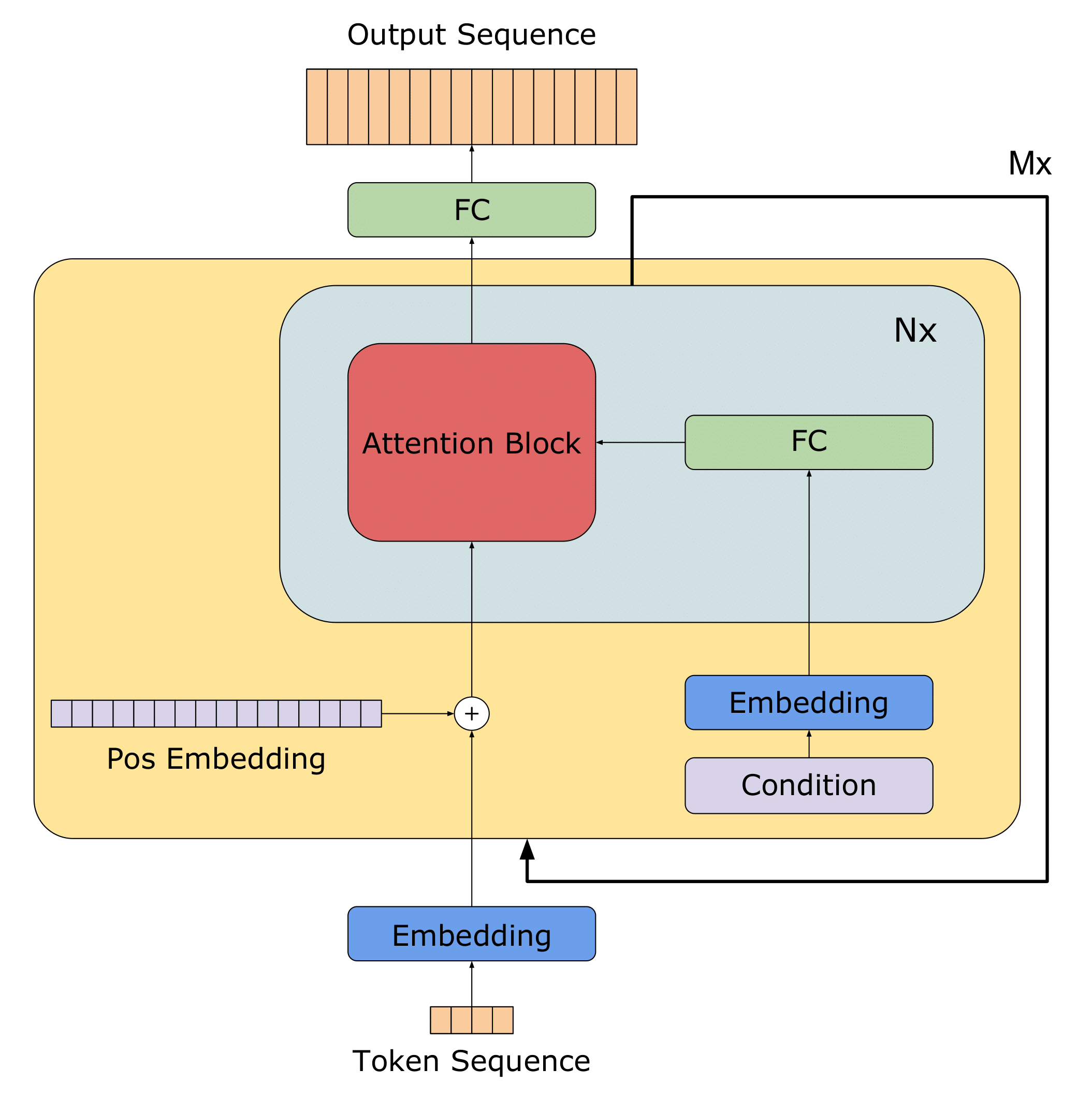}
         \caption{Generator}
         \label{fig:y equals x}
     \end{subfigure}
     \begin{subfigure}[b]{0.35\textwidth}
         \centering
         \includegraphics[width=\textwidth]{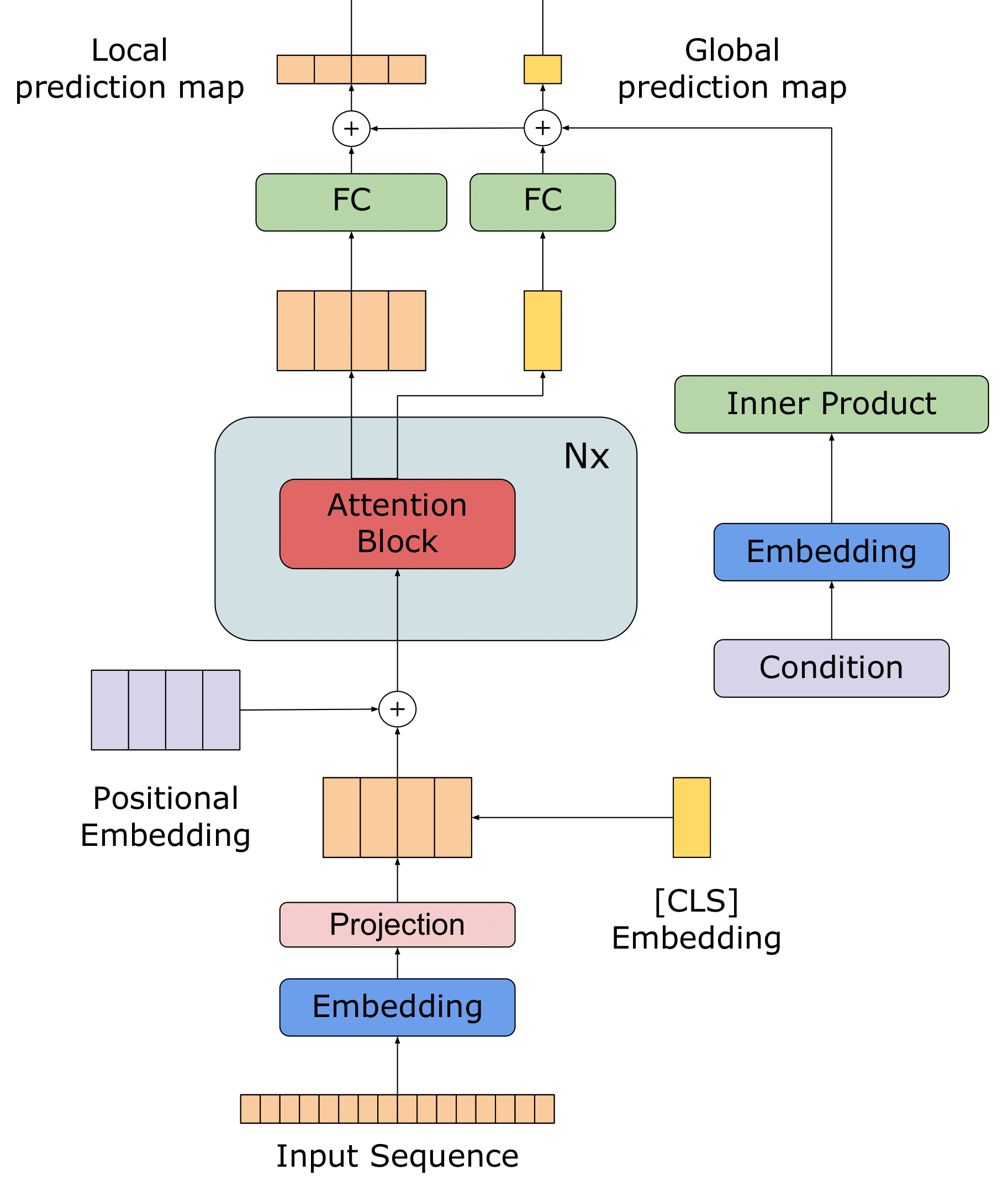}
         \caption{Discriminator}
         \label{fig:three sin x}
     \end{subfigure}
    \caption{Generator and Discriminator. FC stands for Fully-Connected Layer. N is the number of self-attention blocks from which the models are made. M represents the number of autoregressive steps the Generator executes, that is, the sequence length.}
    \label{fig:gen-disc}
\end{figure}

Each of these two outputs serves the purpose of incorporating priors about musical structure into our model. A common way to frame the task of musical generation is as a language modelling one: each individual symbol is treated as a word, and these words compose phrases, periods, and so on. Using this analogy, the goal behind the proposed local loss is to inform if each particular sentence in a text is realistic or not, or, translating that to music, if every short musical idea in the form of a phrase or part of a phrase is realistic or not. The global prediction unit, on the other hand, acts as a signal that encapsulates the overall quality of the sequence and its ability to convey the desired emotional state, complementing the local prediction map. The networks are illustrated in Figure \ref{fig:gen-disc}.



\subsection{Datasets}

We used two datasets to train our models, mainly as a means to allow the networks to use a larger training corpus. Both consist only of songs performed on piano. The AILABS17k dataset \cite{cp} contains over $108$ hours of piano covers of pop songs automatically transcribed by a state-of the art piano transcription model \cite{onsets-and-frames} and converted into MIDI files. The EMOPIA dataset \cite{emopia} was constructed in a similar fashion to the one above, but its songs were afterwards labeled by the authors of the dataset according to perceived sentiment \cite{valence-arousal}. The clips on this dataset amount to approximately $11$ hours. We used  AILABS 17K to allow the networks to learn a general representation of music in a larger corpus, while also being trained on a smaller collection of songs that contains annotated data.


\subsection{Training}

We use the data representation proposed in \cite{remi}, which consists of the NOTE$\_$ON, NOTE$\_$DURATION, VELOCITY, TEMPO, BAR and BEAT events described previously. Before the introduction of the Discriminator, the Generator was trained until convergence through the teacher forcing method ($26000$ steps). This pre-training stage guaranteed the stability of the adversarial stage that was to follow \cite{transformer-gan1, transformer-gan2, relgan}. In this stage, the Generator was trained simultaneously on both datasets, and taking into consideration the difference in size between these datasets, to balance the training process, we alternated between optimization steps on randomly sampled batches from each set. The network worked with sequences of length $2048$, corresponding, on average, to $1$ minute of content.


For the adversarial stage, we used sequences of size $128$. In order to reduce the effects of gradient variance that are inherent to the sampling process, sequences with length $16$ originated from the real set were given to the Generator such that it could have a starting point to perform generation \cite{transformer-gan1}. The Discriminator worked with subsequences of length $16$, and the training was done exclusively on the EMOPIA dataset \cite{emopia}. 

The networks were trained via a combination of the teacher forcing objective plus the RSGAN objective \cite{rsgan} with gradient penalty \cite{wgan-gp}. We performed $1$ optimization step of the Discriminator per Generator step, using a learning rate of $1\cdot10^{-4}$. In total, $26000$ global optimization steps were performed. The overall objective for the generator in this training stage is:

\begin{align}\label{eq:obj_g}
   &\mathcal{L}_{G} = \mathcal{L}_{\text{MLE}} + \alpha\mathcal{L}_{\text{RSGAN}_G\text{-global}} + \beta\mathcal{L}_{\text{RSGAN}_G\text{-local}}, \\
   &\text{where } \mathcal{L}_{mle} = -\mathds{E}_{x \sim P}[\log G_\theta(x)]
\end{align}
where the factors $\mathcal{L}_{\text{RSGAN}_G\text{-global}}$ and $\mathcal{L}_{\text{RSGAN}_G\text{-local}}$ are respectively the global and local GAN losses detailed above, $\alpha$ and $\beta$, which we empirically chose to be equal to $1$, are hyperparameters controlling the relative intensity of each loss factor, and $\mathcal{L}_{\text{MLE}}$ is the Maximum Likelihood. The Discriminator was simply trained with the local and global RSGAN objectives given previously in Equation \ref{eq:rsgan_d} plus the global and local gradient penalties based on Equation \ref{eq:wgan_gp} and regulated by a hyperparameter $\lambda$ (which as per \cite{wgan-gp}, we chose to be 10) . This loss is expressed as Equation \ref{eq:obj_d}. An algorithm outlining the training scheme is available in the supplementary material.

\begin{align}\label{eq:obj_d}
\begin{split}
   \mathcal{L}_{D} = &\mathcal{L}_{\text{RSGAN}_D\text{-global}} + \beta\mathcal{L}_{\text{RSGAN}_D\text{-local}} +\\ &\lambda(\mathcal{L}_{\text{GP-global}} + \mathcal{L}_{\text{GP-local}}).
 \end{split}
\end{align}

\section{Experiments}

We evaluated our models both with respect to the overall quality of the samples they produce and their ability to generate songs that convey the conditioning emotional signals. 
To achieve this purpose, we used both the automatic evaluation metrics proposed in \cite{evaluation-music, musegan} and a set of human evaluation metrics to compare our GAN with the system that, as far as we know, corresponds to a state-of-the-art generative model of symbolic music conditioned by sentiment currently available in the literature. Specifically, our model was compared with the Compound-Word Transformer \cite{cp} variation used by the authors of the EMOPIA article \cite{emopia} to generate music conditioned by emotional class. We also compared our adversarially trained model with a Vanilla Transformer model that was not trained with the adversarial scheme. This model corresponds to the version of the generator network obtained after the pretraining was completed. Code for this work, along with audio samples, are available at \footnote{\href{http://github.com/pneves1051/transformers\_sentiment}{http://github.com/pneves1051/transformers\_sentiment}}.

For the automatic metrics, we chose Pitch Range (distance between highest and lowest pitch), Number of Pitch Classes, and Polyphony (the average number of simultaneous notes). These metrics were calculated using the Muspy library \cite{muspy}. For each model, we generated $400$ samples ($100$ for each class) and evaluated these samples with respect to the characteristics above, then averaged the results to produce the overall model score. The results are presented in Table \ref{tab:results1}.

\begin{table}[!ht]
\centering
\small
\begin{tabular}{|l|lll|}\hline
                                  & PR    & NPC & POLY \\ \hline
Real Data (EMOPIA)\cite{emopia} & $50.94$ & $8.50$ & $5.60$ \\ \hline
Baseline \cite{emopia}            & $49.76$ & $\bm{8.52}$ & $4.36$ \\
Transformer                       & $48.79$ & $8.65$ & $4.37$ \\
Transformer GAN                   & $\bm{50.73}$ & $9.45$ &   $\bm{4.43}$   \\ \hline
\end{tabular}
\caption{Comparison between the samples generated by ours and a state-of-art model. PR stands for Pitch Range, NPC is Number of Pitch Classes and POLY is Polyphony. The best results are highlighted in bold.}
\label{tab:results1}
\end{table}

As it can be seen, our model obtains a superior performance than that of the baseline and the pre-trained model on two of the three metrics. Furthermore, it should be mentioned that the Generator is significantly smaller than the baseline, having only $6$ Attention layers, while the latter has $12$. To be more precise, the baseline has $\sim40M$ parameters, while our generator has $\sim25M$ and our Discriminator has $\sim27M$. Also relevant is the fact that the representation used to train the baseline, that is, the Compound-Word representation \cite{cp}, is more complex than the REMI representation that we use \cite{remi}, and has also been proved to be better than REMI. Since the focus of our work was to implement the GAN framework within the context of symbolic music generation conditioned by sentiment, and given the complexity of this framework, especially when it is applied to the discrete domain, we chose to use a simpler representation in order to maintain the focus of our work on the implementation of the GAN. But as our results show, adapting it to work within the adversarial context, which we leave to future work, could bring about even better results.




Finally, we performed a survey in which participants were asked to judge the musical excerpts generated by the models both in terms of their overall quality and their ability to convey the desired sentiments. Specifically, participants rated the samples on a 5-point Likert scale that ranged from very low to very high with respect to the following characteristics:  Human-likeness, Originality, Structure, Overall Quality, Valence, and Arousal. Details from each characteristic and the corresponding scale are given in the supplementary material. The participants were recruited from the researchers' online circles. Each of the participants had to listen to $12$ musical excerpts in total, $4$ from each model, and within those $4$ item groups, one from each of the $4$ emotional classes. Before the test, some text explaining the basic concepts behind the research was shown to the participants. In total, 18 individuals participated in the experiment.


We present the average participants' scores given to each model for Human-likeness, Originality, Structure and Overall Quality in Table \ref{tab:human-test}. These results suggest that both the proposed Transformer, and the Transformer GAN, are competitive with a state-of-the-art model with respect to the four qualitative metrics.

\begin{table*}[hb]
\centering
\small
\begin{tabular}{l|lllll}
                       & H & O & S & OQ\\ \hline
Baseline \cite{emopia} & 3.32 $\pm$ 1.29  & 2.93 $\pm$ 1.13 & 3.18 $\pm$ 1.30  & 3.49 $\pm$ 1.04   \\
Transformer          & 3.75 $\pm$ 1.24  &  3.22 $\pm$ 1.19  & 3.76 $\pm$ 1.14 & 3.89 $\pm$ 1.14   \\
Transformer-GAN       & 3.56 $\pm$ 1.34  & 3.06 $\pm$ 1.21  & 3.38 $\pm$ 1.09  & 3.44 $\pm$ 1.15  \\ \hline
\end{tabular}
\caption{Results of the Survey where participants were asked to rate the samples generated by several models. The columns are, respectively, Human-Likeness, Originality, Structure and Overall Quality.}
\label{tab:human-test}
\end{table*}

\begin{figure}[ht]
    \centering
    \includegraphics[width=0.45\textwidth]{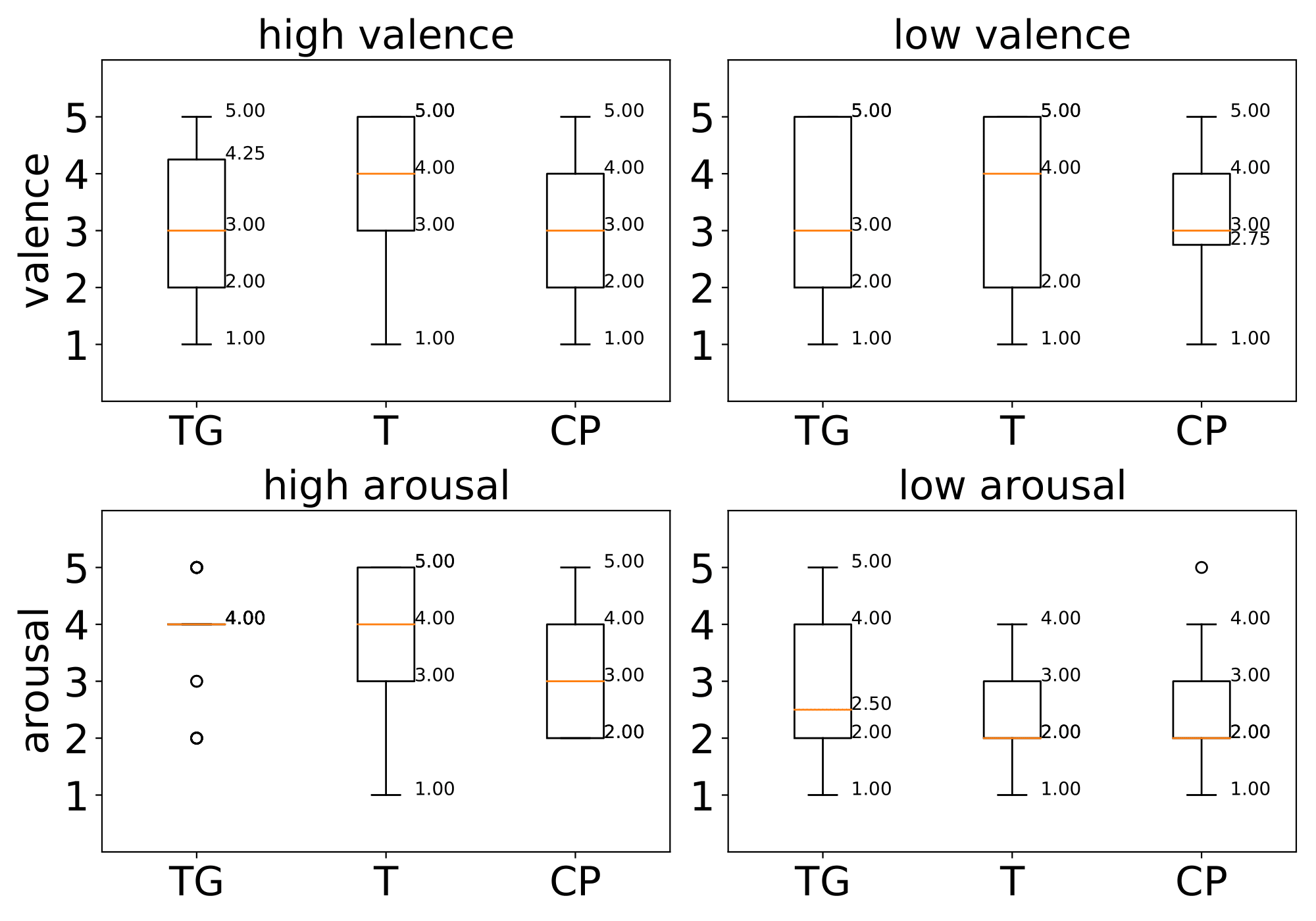}
    \caption{Results of the experiment where participants rated musical samples according to their perceptions about valence and arousal. The acronyms TG, T and CP correspond, respectively, to the Transformer GAN, Transformer and Compound-Word Transformer Baseline models.}
    \label{fig:valence-arousal-result}
\end{figure}

On the next step of the evaluations, we took the participants' answers to the questions related to Valence and Arousal and compared them to the real emotional labels provided to the model during the generation process. Figure \ref{fig:valence-arousal-result} illustrates the results of this experiment.

Once again, we find that the models we developed obtain a performance that competes with that of the current state-of-the-art. By comparing the medians and quartiles of these boxplots, we can draw several conclusions. Firstly, all models seem to have more difficulty capturing valence than they do arousal. This may be an indication  that musical aspects which have a great impact over arousal, such as velocity and tempo, are more easily understood by neural networks, while factors which have an impact over valence, such as modality or frequency of harmonic change \cite{music-and-emotion}, are more difficult to model for these systems.

Furthermore, we see that, in general, our models do not stand behind when compared to the baseline. Between the three, by observing the boxplots, it seems that while the Transformer and the Compound-Word baseline sometimes produce samples that situate themselves more strongly to the side to which they theoretically pertain (e.g., for the high valence and low arousal categories), the Transformer GAN surpasses the simple Transformer due to the fact that it never situates more than $50\%$ of the excerpts on the incorrect side of the middle line, which in this case is represented by the number 3. While the excerpts generated by the Compound Word Transformer also show this same characteristic, we see that there is not a strong reason for us to believe that one stands out in comparison to the other.


Overall, given the superior ratings of the Transformer GAN respective to the automatic metrics and its competitiveness with a state-of-the art model with respect to the human evaluations, and given the considerations above about model size and representation, the Transformer GAN seems to be a promising model for music generation conditioned by sentiment.

\subsection{Conclusion and future work}

We introduced a model capable of generating musical excerpts conditioned by labels that represent perceived emotion. Through the use of MLE pre-training and adversarial training, we guided our model towards understanding some aspects of the relationship between musical structure and affect. Our experiments show that both in terms of quality and the ability to communicate emotion, the samples generated by the proposed model achieve competitive results. Furthermore, our work points to several possible avenues for future research, such as the use of other symbolic representations of music, the development of generative models conditioned by emotion that work directly on audio, and further exploration of the use of affective conditioning signals, e.g., using them to guide automatic composition second-by-second or to automatically generate royalty free music notation based on specific sentiments.

\pagebreak

\bibliography{ISMIRtemplate}

\end{document}